\documentclass{aastex}
\usepackage{spr-astr-addons,natbib}
\usepackage{url}

\RequirePackage{color}

\begin{document}

\title{Exact Scalar-Tensor Cosmological Solutions via Noether Symmetry}
\shorttitle{Exact Scalar-Tensor Cosmological Solutions} %
\shortauthors{Belinch\'{o}n et al.}

\author{J. A. Belinch\'{o}n\altaffilmark{1}} \and \author{T. Harko \altaffilmark{2}} \and \author{M. K. Mak \altaffilmark{3}}
\affil{$^{1}$Dept. F\'{\i}sica, Facultad Ciencias Naturales, Universidad de Atacama,
Copayapu 485, Copiap\'o, Chile, E-mail: abelcal@ciccp.es\\ $^{2}$Department of Mathematics, University College London, Gower Street, London
WC1E 6BT, United Kingdom, E-mail:
t.harko@ucl.ac.uk\\ $^{3}$Dept. F\'{\i}sica, Facultad Ciencias Naturales, Universidad de Atacama,
Copayapu 485, Copiap\'o, Chile, E-mail: mkmak@uda.cl}

\begin{abstract}
In this paper, we investigate the Noether symmetries of a generalized
scalar-tensor, Brans-Dicke type cosmological model, in which we consider
explicit scalar field dependent couplings to the Ricci scalar, and to the
scalar field kinetic energy, respectively. We also include the scalar field
self-interaction potential into the gravitational action. From the condition
of the vanishing of the Lie derivative of the gravitational cosmological
Lagrangian with respect to a given vector field we obtain three cosmological
solutions describing the time evolution of a spatially flat
Friedman-Robertson-Walker Universe filled with a scalar field. The
cosmological properties of the solutions are investigated in detail, and it
is shown that they can describe a large variety of cosmological evolutions,
including models that experience a smooth transition from a decelerating to
an accelerating phase.
\end{abstract}



\section{Introduction}

Despite the fact that general relativity (GR) has passed a variety of
astrophysical and cosmological observational tests, from the weak post-Newtonian regime to the
strong field regime of neutron stars, or even black holes \citep{0}, and
there are some progresses in quantum gravity \citep{qg}, presently it is
generally accepted that there are many shortcomings of GR in both cosmology
and quantum field theory.

In the past twenty years, in view of the various observational evidence
obtained from the study of distant Type Ia Supernovae, it has been showed
that the Universe has undergone a late time accelerated expansion \citep%
{1,2,3,4}. In order to explain this cosmological effect, a mysterious
component, called dark energy, has been proposed for describing the late
time evolution of the Universe. There are many models for dark energy, which
try to explain the observational evidence of the recent acceleration of the
Universe. For example, a canonical scalar field called quintessence was
introduced as a candidate for dark energy \citep{ST}. Scalar fields were
allowed to evolve in the Universe, while having couplings of order of unity
to matter in the context of Chameleon cosmology \citep{JK}. The basic idea of
the so called $k$-essence and dynamical attractors models is based on
evolving scalar fields with non-linear kinetic energy terms in the action
\citep{PJS}. An exotic fluid known as Chaplygin gas was introduced in order
to unify dark matter and dark energy, with the equation of state $p=-A/\rho
^{\alpha }$, where $p$ is the pressure, $\rho $ is the density, $A$ is a
positive constant and $\alpha $ is a constant \citep{5a,6a}. The topics on
phantom dark energy and cosmic doomsday were discussed in \cite{RC}. For
reviews of dark energy models see \citep{7a,8a}. Recently, from the
observational point of view, the Cosmic Microwave Background (CMB) temperature
anisotropies measured by the Wilkinson Microwave Anisotropy Probe and the
Planck satellites were reported in \citep{5,6}, confirming that the total
amount of baryonic matter in the Universe is very low. It is worth to note
that the study of the baryonic acoustic oscillations have also confirmed the
existence of the dark energy, and these results are also correlated with
other cosmological observations \citep{7}.

Another approach to dark energy, related to the modified gravitational
theories, is proposed in order to deal with the cosmic acceleration problem.
In these models the geometric part of Einstein's gravitational field
equation is modified. One of such modifications is the $f\left( R\right) $
gravity theory \citep{9,10,11,12}. Note that Einstein was the first person to
introduce the idea of teleparallel equivalent of GR for the sake of unifying
the gravity and the electromagnetism \citep{13,14}. In teleparallel gravity
theory the gravitational field equations can be obtained by varying the
action with respect to the vierbein fields \citep{15}. 

It is generally recognized that scalar-tensor theory (STT) of gravity is the
most straightforward generalization of standard GR theory. In 1961, the
Brans-Dicke theory was formulated. In this formalism, an additional scalar
field $\phi $ besides the metric tensor $g_{\mu \nu }$ and a dimensionless
coupling constant $\omega $ were introduced in order to describe the
gravitational interaction \citep{BD,B1,B2,B3,B4}. Brans-Dicke theory recovers
the results of GR for large value of the coupling constant $\omega $, that
is for $\omega >500$. Similar formalisms were developed earlier by Jordan
\citep{J1,J2}. One can find the history and developments of STT in \cite%
{h1,h2,h3,h4}. In 1974, the most general STT including four arbitrary
functions of the scalar field and kinetic term was proposed by Horndeski
\cite{ho}. Horndeski's theory can be reduced to various modified gravity
models by choosing the specific four functions. Thus this theory includes a
wide class of modified gravity models, and hence we treat it as an effective
theory of the generalized theories of gravity. A review of the effective
field theory of the inflation/dark energy and the Horndeski's theory was
presented in \cite{K1}. In 2015, several pioneering works beyond Horndeski
were presented in \citep{K2,K3,K4,K5}.

Scalar-tensor theories are promising because they can provide an explanation
of the inflationary behavior in cosmology. Note that the scalar field can be
considered as an additional degree of freedom of the gravitational
interaction. Scalar-tensor theories have been used to model dark energy,
because the scalar fields are good candidates for phantom and quintessence
fields \citep{peeble,TLM,TM1}. Furthermore, it is natural to couple the
scalar field to the curvature after compactification of higher dimensional
theories of gravity such as Kaluza-Klein and string theory, thus offering
the possibility of linking fundamental scalar fields with the nature of the
dark energy. From observational point of view, the parameterized
post-Newtonian (PPN) parameters $\gamma $ and $\beta $ for general STT was
computed, and it was suggested that the PPN parameters $\gamma $ and $\beta $
given by the condition $\left\vert \gamma -1\right\vert \sim $ $\left\vert
\beta -1\right\vert \sim 10^{-6}$ may be detectable by a satellite that
carries a clock with fractional frequency uncertainty $\frac{\Delta f}{f}%
\sim 10^{-16}$ in an eccentric orbit around the Earth \citep{PPN}.

The weak field limit of STT of gravitation was discussed in view of
conformal transformations in \cite{A}, and a new reconstruction method of
STT based on the use of conformal transformations was proposed. This method
allows the derivation of a set of interesting exact cosmological solutions
in Brans-Dicke gravity, as well as in other extensions of GR \citep{S}. The
phase space of Friedmann-Lemaitre-Robertson-Walker models derived from STT
in the presence of the non-minimal coupling $F\left( \phi \right) =\xi \phi
^{2}$ and the effective potential $V\left( \phi \right) =\lambda \phi ^{n}$
was studied in \cite{FS}. Testing the feasibility of scalar-tensor gravity
by scale dependent mass and coupling to matter was investigated in \cite{DF}%
. Recently, an isotropic model in the presence of variable gravitational
and cosmological constants was studied in the context of scalar-tensor
cosmology in \cite{Tony}. Generalized self-similar STT was investigated and
some new exact self-similar solutions were obtained in \cite{Tony1}, by
using the Kantowski-Sach models.

A powerful mathematical method that allows to handle cosmological models coming from different fundamental gravity theories is represented by the Noether symmetry approach \citep{Capo96}. With its use one can solve exactly the cosmological equations in several gravitational theories. The interest in such a mathematical approach becomes greater when the solutions found are physically interesting. In \cite{Capo96} several cosmological solutions obtained by using Noether symmetries are presented. They are Friedmann, power law, pole-like and de-Sitter-like, and thus they can cover a large range of expected cosmological behaviors. From the observational point of view such Noether symmetrical solutions can be interpreted as the background on which one can compare theoretical models with the observational data from large-scale structure formation and Cosmic Microwave Background Radiation. These background solutions can be also used to formulate the theories of cosmological perturbations for a given model.

A complete Noether symmetry analysis in the framework of scalar-tensor
cosmology was reported in \cite{NS} (see also \citep{MNS,S1} for other
approaches). Scalar-tensor cosmology with $1/R$ curvature correction by
Noether symmetry was discussed in \cite{NN}. Recently, Noether symmetries of
some homogeneous Universe models in curvature corrected scalar-tensor
gravity was studied in \cite{NS1}. Using Noether symmetry approach, the
solutions of the gravitational field equations in the context of scalar
tensor teleparallel dark gravity was obtained in \cite{YK}. The evolution of
teleparallel dark energy cosmological model with a fermionic field was
studied in \cite{YK1}. The evolution of a three-dimensional minisuperspace
cosmological model via Noether symmetry was investigated in \cite{BV}. With
the help of Noether symmetry, some cosmological analytical solutions for
specific functional forms of coupling and potential functions were obtained
in \cite{AP}. The application of point symmetries in the recently proposed
metric-Palatini hybrid gravity in order to select the $f\left( R^{\ast
}\right) $ functional form and to find analytical solutions for the
gravitational field equations and for the related Wheeler-DeWitt equation
was considered in \cite{AB}, where $R^{\ast }$ is the Palatini curvature
scalar.

With the help of the standard mathematical procedure, the Lie group
approach, and dynamical Noether symmetry techniques, we obtain several exact
solutions of the gravitational field equations describing the time
evolutions of a spatially flat Friedman-Robertson-Walker (FRW) Universe in
the context of the scalar-tensor gravity. The obtained solutions can
describe both accelerating and decelerating phases during the cosmological
expansion of the Universe \citep{submitted}.

It is the purpose of the present paper to investigate the Noether symmetries
of the cosmological dynamics of a modified STT gravitational model \citep%
{n1,n2,n3}, in which there is an explicit scalar field dependent coupling to the
Ricci scalar, and to the scalar field kinetic energy, respectively are
included in the gravitational action. In this model the scalar field
self-interaction potential is also explicitly considered. From the condition
of the vanishing of the Lie derivative of the gravitational cosmological
Lagrangian with respect to a given vector field we obtain three cosmological
solutions, describing the time evolution of a spatially flat FRW Universe
filled with a scalar field. The cosmological properties of these solutions
are investigated in detail, and it is shown that they can describe a large
variety of cosmological evolutions, including models that experience a
smooth transition from a decelerating to an accelerating phase. Thus the
presented solutions may offer a possible explanations of the observed
cosmological dynamics, without the need of considering dark energy.

The present paper is organized as follows. We formulate the scalar-tensor
model of gravity in Section~\ref{sect1}. The Noether symmetries of this STT
model are investigated, and two general analytical solutions with six
particular solutions of the symmetry equation are presented in Section~\ref%
{sect2}. The cosmological implications of three solutions are investigated
in detail in Section~\ref{sect3}. Finally, we discuss and conclude our
results in Section~\ref{sect4}.

\section{Formalism of scalar-tensor gravity}\label{sect1}

In the present paper we consider a generalized scalar-tensor gravitational
model, in which the Ricci scalar $R$, describing pure gravity, couples in a
non-standard way with a scalar field $\ \phi $. Moreover, we assume the
second non-standard coupling between the scalar field and the kinetic energy
term. Hence for this model the gravitational action takes the form \citep%
{n1,n2,n3}
\begin{equation}
S=\int d^{4}x\sqrt{-g}\left[ F\left( \phi \right) R+\frac{1}{2}Z\left( \phi
\right) g^{\mu \nu }\phi _{;\mu }\phi _{;\nu }-V\left( \phi \right) \right] ,
\label{Act}
\end{equation}%
where $F\left( \phi \right) $ is the generic function describing the
coupling between the scalar field and geometry, $Z\left( \phi \right) $ is
the arbitrary function coupling the scalar field to its kinetic energy, also
distinguishing between various STT of gravity, and $V\left( \phi \right) $
is the scalar field potential.

In the following we investigate only the cosmological implications of this
model. We consider the case of a spatially flat FRW Universe, described by
the line element
\begin{equation}
ds^{2}=dt^{2}-a^{2}(t)\left( dx^{2}+dy^{2}+dz^{2}\right) ,  \label{1}
\end{equation}%
where $a(t)$ is the scale factor, which gives the information on the
expansion of the universe. With the help of the line element (\ref{1}), we
obtain the Ricci scalar as
\begin{equation}
R=-6\left[ \frac{\ddot{a}}{a}+\left( \frac{\dot{a}}{a}\right) ^{2}\right] ,
\label{R}
\end{equation}%
where the overdot denotes the derivative with respect to the time $t$. The
Lagrangian density on the configuration space $\left( a,\phi \right) $ for
the non-minimal coupled scalar-tensor cosmology described by action (\ref%
{Act}) takes the form%
\begin{eqnarray}  \label{L}
\mathcal{L} &=&6a\dot{a}^{2}F(\phi )+6a^{2}\dot{a}\dot{\phi}F^{\prime }(\phi
)+ a^{3}\Bigg[ \frac{1}{2}Z(\phi )\dot{\phi}^{2}-V(\phi )\Bigg]  \nonumber \\
&=&a^{3}\Bigg[ 6\left( \frac{\dot{a}}{a}\right) ^{2}F)\phi )+6\frac{\dot{a}}{%
a}\dot{F}(\phi )+\frac{Z(\phi )\dot{\phi}^{2}}{2}-V(\phi )\Bigg] ,  \nonumber
\\
\end{eqnarray}%
where the prime denotes the derivative with respect to the scalar field $%
\phi $. A first class of solutions of the model can be obtained by assuming
that the Lagrangian $\mathcal{L}$ is degenerate. Then, the Hessian $\mathcal{%
H}$ defined mathematically as%
\begin{equation}
\mathcal{H}=\left\vert
\begin{array}{cc}
\frac{\partial ^{2}\mathcal{L}}{\partial \dot{a}^{2}} & \frac{\partial ^{2}%
\mathcal{L}}{\partial \dot{a}\partial \dot{\phi}} \\
\frac{\partial ^{2}\mathcal{L}}{\partial \dot{\phi}\partial \dot{a}} & \frac{%
\partial ^{2}\mathcal{L}}{\partial \dot{\phi}^{2}}%
\end{array}%
\right\vert ,  \label{2a}
\end{equation}%
vanishes, $\mathcal{H}=0$. By inserting Eq. (\ref{L}) into Eq. (\ref{2a}),
and using the condition for degeneracy $\mathcal{H}=0$, the latter condition
gives
\begin{eqnarray}  \label{Hess}
\left\vert
\begin{array}{cc}
12aF & 6a^{2}\frac{dF}{d\phi } \\
6a^{2}\frac{dF}{d\phi } & a^{3}Z%
\end{array}%
\right\vert &=&12a^{4}\left[ Z(\phi )F(\phi )-3\left( \frac{dF(\phi )}{d\phi
}\right) ^{2}\right]  \nonumber \\
&=&0.
\end{eqnarray}

Immediately Eq. (\ref{Hess}) can be integrated to yield the integral equation%
\begin{equation}
2\sqrt{F\left( \phi \right) }+\frac{\sqrt{3}}{3}\int \sqrt{Z(\phi )}d\phi
\pm C=0,  \label{qq}
\end{equation}%
where $C$ is an arbitrary constant of integration. By choosing some specific
forms for one of the arbitrary functions $F\left( \phi \right) $ or $Z(\phi
) $ we can fix the coupling functions of the model.

For example, if $F\left( \phi \right) =\mu ^{2}\phi ^{2s}/4$, with $\mu $
and $s$ arbitrary constants, from Eq.~(\ref{Hess}) then we obtain $Z(\phi
)=3\mu ^{2}s^{2}\phi ^{2s-2}$. By assuming for $Z(\phi )$ the functional
form $Z(\phi )=3m_{1}^{2}Z_{1}^{2}\phi ^{2\left( m_{1}-1\right) }$, where $%
m_{1}$ and $Z_{1}$ are arbitrary constants, we obtain for $F(\phi )$ the
expression $F(\phi )=\left( C-Z_{1}\phi ^{m_{1}}\right) ^{2}/4$, where we
have adopted the minus sign for the constant $C$ in Eq.~(\ref{qq}).

If we fix both $F\left( \phi \right) $ and $Z(\phi )$ as having a polynomial
form in $\phi $, we obtain an algebraic equation which has the solution
given by $\phi $ = constant. In this case the physical model described by
the degenerate Lagrangian (\ref{L}) reduces, after a rescaling of the
parameters, to standard GR, with the potential $V(\phi )$ generating a
cosmological constant. Note that the conditions $\frac{\partial \mathcal{L}}{%
\partial t}=0$ and $\mathcal{H}\neq 0$ hold for the time-independent,
non-degenerate Lagrangians $\mathcal{L=L}\left( q^{i},\dot{q}^{j}\right) $
\cite{C}.

Next, in order to obtain some cosmological solutions describing the time
evolution of the FRW Universe in the context of the cosmological model
described by the action (\ref{Act}), we will first investigate in the
following Section its Noether symmetries.

\section{Noether symmetries}

\label{sect2}

We define the configuration space $Q$ of the cosmological model as $Q\equiv
(a,\phi )$, with the tangent space $TQ$ given by $TQ\equiv (a,\dot{a},\phi ,%
\dot{\phi})$. Using the infinitesimal generator of the Noether symmetry, the
lift vector $\mathbf{X}$ can now be written as
\begin{equation}
\mathbf{X}=\alpha \left( a,\phi \right) \frac{\partial }{\partial a}+\beta
\left( a,\phi \right) \frac{\partial }{\partial \phi }+\frac{d\alpha }{dt}%
\frac{\partial }{\partial \dot{a}}+\frac{d\beta }{dt}\frac{\partial }{%
\partial \dot{\phi}},  \label{X}
\end{equation}%
where $\alpha ,\beta $ are functions of the scale factor $a$ and the scalar
field $\phi $. The existence of the Noether symmetry implies the existence
of a vector field $\mathbf{X}$, so that we have the relation \citep{C}
\begin{equation}
L_{\mathbf{X}}\mathcal{L}=\mathbf{X}\mathcal{L}=0,  \label{LX}
\end{equation}%
where $L_{\mathbf{X}}$ stands for the Lie derivative with respect to the
vector field $\mathbf{X}$. The existence of the symmetry $\mathbf{X}$
provides us a constant of motion, through the Noether theorem. The constant
of motion $\Sigma $, defined by the inner derivative, takes the form
\begin{equation}
\Sigma =i_{\mathbf{X}}\theta _{\mathcal{L}},  \label{i1}
\end{equation}
where the Cartan one form is defined by
\begin{equation}
\theta _{\mathcal{L}}=\frac{\partial \mathcal{L}}{\partial \dot{a}}da+\frac{%
\partial \mathcal{L}}{\partial \dot{\phi}}d\phi .  \label{i2}
\end{equation}%
Therefore, with the help of Eqs.~(\ref{L}) and (\ref{LX}), we obtain the
following system of partial differential equations%
\begin{equation}
F\left( \alpha +2a\partial _{a}\alpha \right) +aF^{\prime }\left( \beta
+a\partial _{a}\beta \right) =0,  \label{x1}
\end{equation}%
\begin{equation}
3Z\alpha +Z^{\prime }a\beta +12F^{\prime }\partial _{\phi }\alpha
+2Za\partial _{\phi }\beta =0,  \label{x2}
\end{equation}%
\begin{equation}
a\beta F^{\prime \prime }+\left( 2\alpha +a\partial _{a}\alpha +a\partial
_{\phi }\beta \right) F^{\prime }+2F\partial _{\phi }\alpha +\frac{1}{6}%
Za^{2}\partial _{a}\beta =0,  \label{x3}
\end{equation}%
\begin{equation}
3\alpha V+a\beta V^{\prime }=0.  \label{x4}
\end{equation}

In the following we shall obtain two classes of solutions of the system (\ref%
{x1})-(\ref{x4}), and we shall investigate their cosmological properties.

\subsection{Solution 1}

It is easy to show by direct computations that Eqs.~(\ref{x1})-(\ref{x4})
have the following solution,
\begin{eqnarray}  \label{xx}
F\left( \phi \right) &=&K\phi ^{n}, Z\left( \phi \right) =Z_{0}\phi ^{n-2},
V\left( \phi \right) =V_{0}\phi ^{n},  \nonumber \\
&&\alpha \left(a\right) =c_{1}a, \beta \left( \phi \right) =-\frac{3c_{1}}{n}%
\phi ,
\end{eqnarray}
where $K$, $Z_{0}$, $V_{0}$, $n$ and $c_{1}$ are the arbitrary constants.

\subsection{Solution 2}

After tedious calculations, we integrate Eqs.~(\ref{x1})-(\ref{x4}) to
obtain a second class of solutions as
\begin{equation}
F\left( \phi \right) =K\phi ^{n}, V\left( \phi \right) =V_{0}\phi ^{\frac{%
3c_{1}n}{2c_{1}+1}},
\end{equation}
\begin{equation}
Z\left( \phi \right) =\frac{3Kn^{2}\left[ \frac{4c_{2}c_{1}\left(
c_{1}+1\right) }{\left( 2c_{1}+1\right) ^{2}}\phi ^{\frac{3c_{1}n}{2c_{1}+1}%
}-c_{3}\phi ^{n}\right] }{\left( c_{2}\phi ^{\frac{\left( c_{1}-1\right) n}{%
2c_{1}+1}}-c_{3}\right) \phi ^{2}},  \label{s1}
\end{equation}%
\begin{eqnarray}  \label{s2}
&&\alpha \left( a,\phi \right) =\frac{a^{c_{1}}}{n\left( 1-c_{1}\right) }%
\times  \nonumber \\
&&\sqrt{2n\left( 2c_{1}+1\right) \left( 1-c_{1}\right) \left[ c_{2}\phi ^{%
\frac{2\left( c_{1}^{2}-1\right) n}{2c_{1}+1}}-c_{3}\phi ^{n\left(
c_{1}-1\right) }\right] },  \nonumber \\
\end{eqnarray}
\begin{eqnarray}
\beta \left( a,\phi \right) &=&\frac{1}{c_{1}}\sqrt{\frac{2}{1-c_{1}}\left(
\frac{2c_{1}+1}{n}\right) ^{3}}\times  \nonumber \\
&&\frac{a^{c_{1}-1}\phi \left( c_{3}-c_{2}\phi ^{\frac{\left( c_{1}-1\right)
n}{2c_{1}+1}}\right) }{\sqrt{c_{2}\phi ^{\frac{2c_{1}\left( 1-c_{1}\right) n%
}{2c_{1}+1}}-c_{3}\phi ^{n\left( 1-c_{1}\right) }}},  \label{s3}
\end{eqnarray}
where the arbitrary constant $c_{1}$ satisfies the relations $c_{1}\neq 1$
and $c_{1}\neq -1/2$ , and $c_{2}$ and $c_{3}$ are arbitrary constants of
integration.

In order to investigate in detail the cosmological implications of the
present solutions, in the following we shall concentrate on two particular
solutions of Eqs.~(\ref{s1})-(\ref{s3}), corresponding to $c_{2}=0$ and $%
c_{3}=0$, respectively.

\subsubsection{Specific solution $c_{2}=0$ and $n\neq 0$}

Assume that the arbitrary constant $c_{2}$ vanishes, then from Eqs. (\ref{s1}%
)- (\ref{s3}),
\begin{equation}
F\left( \phi \right) =K\phi ^{n}, V\left( \phi \right) =V_{0}\phi ^{\frac{%
3c_{1}n}{2c_{1}+1}},Z\left( \phi \right) =3Kn^{2}\phi ^{n-2},  \label{sa}
\end{equation}%
\begin{align}
\alpha \left( a,\phi \right) & =\frac{\sqrt{2n\left( 2c_{1}+1\right) \left(
c_{1}-1\right) c_{3}}}{n\left( 1-c_{1}\right) }a^{c_{1}}\phi ^{n\left(
c_{1}-1\right) /2},  \label{sb} \\
\beta \left( a,\phi \right) & =\frac{1}{c_{1}}\sqrt{\frac{2c_{3}}{c_{1}-1}%
\left( \frac{2c_{1}+1}{n}\right) ^{3}}a^{c_{1}-1}\phi ^{1-\frac{n\left(
1-c_{1}\right) }{2}}.  \label{sc}
\end{align}

Now, we shall present two specific solutions of Eqs.~(\ref{sa})-(\ref{sc})
for $n=1,2$ respectively.

For $n=2$, from Eqs. (\ref{sa})-(\ref{sc}), we obtain the following solution
for the generalized STT model,
\begin{equation}
F\left( \phi \right) =K\phi ^{2}, V\left( \phi \right) =V_{0}\phi ^{\frac{%
6c_{1}}{2c_{1}+1}}, Z\left( \phi \right) =12K,  \label{s6}
\end{equation}%
\begin{align}
\alpha \left( a,\phi \right) & =\frac{\sqrt{\left( 2c_{1}+1\right) \left(
c_{1}-1\right) c_{3}}}{\left( 1-c_{1}\right) }a^{c_{1}}\phi ^{c_{1}-1},
\label{av} \\
\beta \left( a,\phi \right) & =\frac{1}{2c_{1}}\sqrt{\frac{c_{3}\left(
2c_{1}+1\right) ^{3}}{c_{1}-1}}a^{c_{1}-1}\phi ^{c_{1}}.  \label{ax}
\end{align}

For $n=1$, we obtain the Jordan-Brans-Dicke type solutions,%
\begin{equation}
F\left( \phi \right) =K\phi , V\left( \phi \right) =V_{0}\phi ^{\frac{3c_{1}%
}{2c_{1}+1}}, Z\left( \phi \right) =3K\phi ^{-1},  \label{s7}
\end{equation}%
\begin{align}
\alpha \left( a,\phi \right) & =\frac{\sqrt{2\left( 2c_{1}+1\right) \left(
c_{1}-1\right) c_{3}}}{\left( 1-c_{1}\right) }a^{c_{1}}\phi ^{\left(
c_{1}-1\right) /2},  \label{n} \\
\beta \left( a,\phi \right) & =\frac{1}{c_{1}}\sqrt{\frac{2c_{3}\left(
2c_{1}+1\right) ^{3}}{c_{1}-1}}a^{c_{1}-1}\phi ^{\frac{1+c_{1}}{2}}.
\label{m}
\end{align}

\subsubsection{Specific solution $c_{3}=0$ and $n\neq 0$}

Assume that the arbitrary constant $c_{3}$ vanishes, then from Eqs. (\ref{s1}%
)-(\ref{s3}), we obtain the solutions%
\begin{eqnarray}  \label{s8}
F\left( \phi \right) &=&K\phi ^{n}, V\left( \phi \right) =V_{0}\phi ^{\frac{%
3c_{1}n}{2c_{1}+1}},  \nonumber \\
Z\left( \phi \right) &=&\frac{12Kn^{2}c_{1}\left( c_{1}+1\right) }{\left(
2c_{1}+1\right) ^{2}}\phi ^{n-2},
\end{eqnarray}
\begin{equation}  \label{s9}
\alpha \left( a,\phi \right) =\sqrt{\frac{2\left( 2c_{1}+1\right) c_{2}}{%
n\left( 1-c_{1}\right) }}a^{c_{1}}\phi ^{\frac{n\left( c_{1}^{2}-1\right) }{%
2c_{1}+1}},
\end{equation}
\begin{equation}  \label{s10}
\beta \left( a,\phi \right) =-\frac{1}{c_{1}}\sqrt{\frac{2c_{2}}{1-c_{1}}%
\left( \frac{2c_{1}+1}{n}\right) ^{3}}a^{c_{1}-1}\phi ^{\frac{n\left(
c_{1}^{2}-1\right) }{2c_{1}+1}+1}.
\end{equation}

Assume that $n=2$ and $c_{1}=-2$, then from Eqs. (\ref{s8})-(\ref{s10}), we
obtain the results
\begin{eqnarray}
F\left( \phi \right) &=&K\phi ^{2}, V\left( \phi \right) =V_{0}\phi ^{4},
Z\left( \phi \right) =\frac{32K}{3},  \nonumber \\
\alpha \left( a,\phi \right) &=&\frac{\sqrt{-c_{2}}}{\left( a\phi \right)
^{2}}, \beta \left( a,\phi \right) =\frac{3\sqrt{-c_{2}}}{4a^{3}\phi },
c_{2}<0,  \label{11}
\end{eqnarray}
which are similar to the solutions given by \cite%
{Capo94}. Furthermore for $n=2$ and $c_{1}=1/4$, we obtain the solutions%
\begin{eqnarray}  \label{12}
F\left( \phi \right) &=&K\phi ^{2}, V\left( \phi \right) =V_{0}\phi ,
Z\left( \phi \right) =\frac{20K}{3},  \nonumber \\
\alpha \left( a,\phi \right) &=&\sqrt{2c_{2}}\left( a\phi ^{-5}\right)
^{1/4}, \\
\beta \left( a,\phi \right)& =&-\frac{3\sqrt{2c_{2}}}{\left( a^{3}\phi
\right) ^{1/4}}, c_{2}>0. \nonumber
\end{eqnarray}

Hence by using the requirement of the existence of a Noether symmetry, and
with the help of the vector field $\mathbf{X}$ given by Eq. (\ref{X}), of
the Lagrangian density (\ref{L}), we have obtained some solutions of the
system (\ref{x1})-(\ref{x4}), in the FRW geometry (\ref{1}). In the
following Sections we shall investigate the cosmological solutions in the
framework of the general STT corresponding to these Noether symmetries.

\section{Exact cosmological solutions}\label{sect3}

\subsection{Cosmological solution 1}

Using the Noether symmetries presented in Section III, we have found a set
of solutions of the symmetry equations, given by Eqs.~(\ref{xx}). Now in
order to find the explicit forms of the scale factor $a\left( t\right) $ and
of the scalar field $\phi \left( t\right) $, we introduce two arbitrary
functions $z$ and $w$ defined as
\begin{equation}
z=z(a,\phi ),\qquad w=w(a,\phi ),  \label{as}
\end{equation}%
respectively. The transformed Lagrangian is cyclic in one of the new
variables. Using the relations $i_{\mathbf{X}}\left( dz\right) =1$ and $i_{%
\mathbf{X}}\left( dw\right) =0$, we obtain the differential equations
\begin{align}
\qquad \alpha \partial _{a}z+\beta \partial _{\phi }z& =1,  \label{v} \\
\alpha \partial _{a}w+\beta \partial _{\phi }w& =0,  \label{w}
\end{align}%
respectively. Now in Eqs.~(\ref{xx}) 
without loss of generality, we set $c_{1}=1$ in the following. By
substituting Eqs.~(\ref{xx}) into Eqs.~(\ref{v}) and (\ref{w}), the latter
equations can respectively be integrated to yield
\begin{align}
z\left( a,\phi \right) & =\ln a+\phi a^{3/n},\qquad w\left( a,\phi \right)
=\phi a^{3/n},  \label{w1} \\
a\left( w,z\right) & =e^{z-w},\qquad \phi \left( w,z\right) =we^{\frac{3}{n}%
\left( w-z\right) }.  \label{w2}
\end{align}%
With the help of Eqs. (\ref{L}), (\ref{xx}), (\ref{w1}), and (\ref{w2}), and
by setting $n\neq 0$, $Z_{0}=1,$ and $8n^{2}K=3$, the Lagrangian (\ref{L})
takes the form%
\begin{equation}
\mathcal{L}=\frac{3}{4n}\dot{w}w^{n-1}\left( \dot{w}-\dot{z}\right) +\frac{1%
}{2}w^{n-2}\dot{w}^{2}-V_{0}w^{n}.  \label{m1}
\end{equation}%
Next the Euler-Lagrange equations are given by
\begin{equation}
\frac{d}{dt}\left( \frac{\partial \mathcal{L}}{\partial \dot{z}}\right) =%
\frac{\partial \mathcal{L}}{\partial z},  \label{m2}
\end{equation}%
\begin{equation}
\frac{d}{dt}\left( \frac{\partial \mathcal{L}}{\partial \dot{w}}\right) =%
\frac{\partial \mathcal{L}}{\partial w},  \label{m3}
\end{equation}%
respectively. By inserting Eq. (\ref{m1}) into Eqs. (\ref{m2}) and (\ref{m3}%
), the latter equations yield
\begin{equation}
\frac{d}{dt}\left( w^{n-1}\dot{w}\right) =0,  \label{m4}
\end{equation}
\begin{eqnarray}
\ddot{z}&=&2\left( 1+\frac{2n}{3w}\right) \ddot{w}+\frac{2}{3}n\left(
n-2\right) \left( \frac{\dot{w}}{w}\right) ^{2}+  \nonumber \\
&&\left( n-1\right) \frac{\dot{w}^{2}}{w}+\frac{4}{3}V_{0}n^{2},  \label{m5}
\end{eqnarray}%
respectively. Eq. (\ref{m4}) can be easily integrated to give%
\begin{equation}
w\left( t\right) =\left[ n\left( k_{1}t+k_{2}\right) \right] ^{1/n},
\label{m6}
\end{equation}%
where $k_{1}$ and $k_{2}$ are two arbitrary constants of integration. By
inserting Eq.~(\ref{m6}) into Eq.~(\ref{m5}), the latter can be integrated
to give
\begin{equation}
z\left( t\right) =\left[ n\left( k_{1}t+k_{2}\right) \right] ^{1/n}+f\left(
t\right) +\frac{2}{3}\ln \left\vert k_{1}t+k_{2}\right\vert +k_{4},
\label{m9}
\end{equation}%
where for simplicity we have denoted the arbitrary function $f\left(
t\right) $ as
\begin{equation}
f\left( t\right) =t\left[ \frac{2}{3}n^{2}V_{0}\left( t-\frac{2k_{2}}{k_{1}}%
\right) +k_{3}\right] ,
\end{equation}%
where $k_{3}$ and $k_{4}$ are arbitrary constants of integration. Now by
inserting Eqs.~(\ref{m6}) and (\ref{m9}) into Eq.~(\ref{w2}), this allows us
to obtain the scale factor $a$, the Hubble function $H=\dot{a}/a$ and the
scalar field $\phi $ as
\begin{align}
a\left( t\right) & =e^{k_{4}}\left( k_{1}t+k_{2}\right) ^{2/3}e^{f\left(
t\right) },  \label{m11} \\
H(t)& =\frac{1}{3}\left[ \frac{4n^{2}V_{0}(k_{1}t-k_{2})}{k_{1}}+\frac{2k_{1}%
}{k_{1}t+k_{2}}+3k_{3}\right] ,  \label{H1} \\
\phi ^{n}\left( t\right) & =\frac{n}{e^{3k_{4}}}\frac{1}{\left(
k_{1}t+k_{2}\right) e^{3f\left( t\right) }},  \label{m10}
\end{align}%
respectively.

By inserting Eq.~(\ref{m10}) into the relation given by $\ V\left( \phi
\right) =V_{0}\phi ^{n}$, thus we easily obtain the potential%
\begin{equation}
V\left( t\right) =\frac{nV_{0}}{e^{3k_{4}}}\frac{1}{\left(
k_{1}t+k_{2}\right) e^{3f\left( t\right) }}.
\end{equation}

An important observational quantity is the deceleration parameter $q$
defined as%
\begin{equation}
q\left( t\right) =\frac{d}{dt}\left( \frac{1}{H}\right) -1.  \label{m12}
\end{equation}

The sign of the deceleration parameter indicates whether the cosmological
model accelerates, or not. The positive sign of $q$ corresponds to standard
decelerating models whereas the negative sign indicates acceleration. Now by
substituting Eq.~(\ref{m11}) into Eq.~(\ref{m12}), the latter becomes%
\begin{equation}
q\left( t\right) =\frac{6\left[ \frac{k_{1}^{2}}{\left( k_{1}t+k_{2}\right)
^{2}}-2n^{2}V_{0}\right] }{\left[ 4n^{2}V_{0}\left( t-\frac{k_{2}}{k_{1}}%
\right) +\frac{2k_{1}}{\left( k_{1}t+k_{2}\right) }+3k_{3}\right] ^{2}}-1.
\end{equation}

It is important to note that for $t\rightarrow 0$ both the scalar field and
its potential are finite if $k_{2}\neq 0$. In order to determine the values
of the integration constants $k_{i}$, $i=1,...,4$ we proceed as follows. We
assume first that at $t=0$, $a\left( 0\right) =a_{0}$, a condition which
gives, with the use of Eq.~(\ref{m11}) $a_{0}=e^{k_{4}}k_{2}^{2/3}$, or $%
k_{2}=a_{0}^{3/2}$. Hence we can take $k_{4}=0$ without any loss of
generality. The condition $H(0)=H_{0}$ gives
\begin{equation}
H_{0}=-\frac{4n^{2}V_{0}k_{2}}{3k_{1}}+\frac{2k_{1}}{3k_{2}}+k_{3}.
\end{equation}%
Again, without any loss of generality we can fix the constant $k_{3}$ as $%
k_{3}=0$. Then from the above equation we obtain for the ratio of $%
k_{1}/k_{2}$ given by
\begin{equation}
\frac{k_{1}}{k_{2}}=\frac{3H_{0}+\sqrt{9H_{0}^{2}+32n^{2}V_{0}}}{4}=\frac{%
3H_{0}}{4}\left( 1+\sqrt{1+\chi }\right) ,
\end{equation}%
where we have denoted $\chi =32n^{2}V_{0}/9H_{0}^{2}$. Therefore the
cosmological parameters corresponding to the first Noether symmetry solution
can be written in the form
\begin{eqnarray}
a(t) &=&a_{0}\left[ \frac{3}{4}\left( 1+\sqrt{1+\chi }\right) H_{0}t+1\right]
^{2/3}\times   \nonumber \\
&&e^{\frac{\chi }{16}H_{0}t\left( 3H_{0}t-\frac{8}{1+\sqrt{1+\chi }}\right)
},
\end{eqnarray}%
\begin{eqnarray}
H(t) &=&H_{0}\Bigg[\frac{\chi }{8}\left( 3H_{0}t-\frac{4}{1+\sqrt{1+\chi }}%
\right) +  \nonumber \\
&&\frac{2\left( 1+\sqrt{1+\chi }\right) }{3\left( 1+\sqrt{1+\chi }\right)
H_{0}t+4}\Bigg],
\end{eqnarray}%
\begin{equation}
\phi ^{n}(t)=\frac{4n}{a_{0}^{3/2}}\frac{e^{\frac{3\chi }{16}H_{0}t\left(
\frac{8}{1+\sqrt{1+\chi }}-3H_{0}t\right) }}{\left[ 3\left( 1+\sqrt{1+\chi }%
\right) H_{0}t+4\right] },
\end{equation}%
\begin{equation}
V(t)=\frac{4nV_{0}}{a_{0}^{3/2}}\frac{e^{\frac{3\chi }{16}H_{0}t\left( \frac{%
8}{1+\sqrt{1+\chi }}-3H_{0}t\right) }}{\left[ 3\left( 1+\sqrt{1+\chi }%
\right) H_{0}t+4\right] },
\end{equation}%
\begin{eqnarray}
q(t) &=&-24\left( \sqrt{\chi +1}+1\right) ^{2}\Bigg[ 9\chi \left( \chi +2%
\sqrt{\chi +1}+2\right) \times \nonumber\\
&&\left( H_{0}t\right) ^{2}+32\left( \sqrt{\chi +1}%
+1\right) \Bigg] ^{-2}
\Bigg\{ 3\chi \Bigg[ 3H_{0}t\nonumber\\
&&\left( \chi +2\sqrt{\chi +1}+2\right) +8\left(
\sqrt{\chi +1}+1\right) \Bigg] H_{0}t \nonumber\\
&&-32\left( \sqrt{\chi +1}+1\right)
\Bigg\} -1.
\end{eqnarray}

\begin{figure}[h]
\centering
\includegraphics[width=0.45\textwidth]{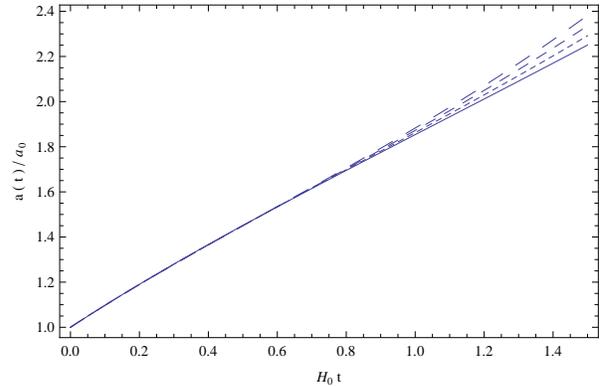}
\caption{ Time variation of the scale factor $a(t)$ in the first Noether
symmetry solution of the generalized STT cosmological model for different
values of the parameter $\protect\chi $: $\protect\chi =0.15$ (solid curve),
$\protect\chi =0.25$ (dotted curve), $\protect\chi =0.35$ (dashed curve),
and $\protect\chi =0.45$ (long dashed curve), respectively.}
\label{fig1}
\end{figure}
\begin{figure}[h]
\centering
\includegraphics[width=0.45\textwidth]{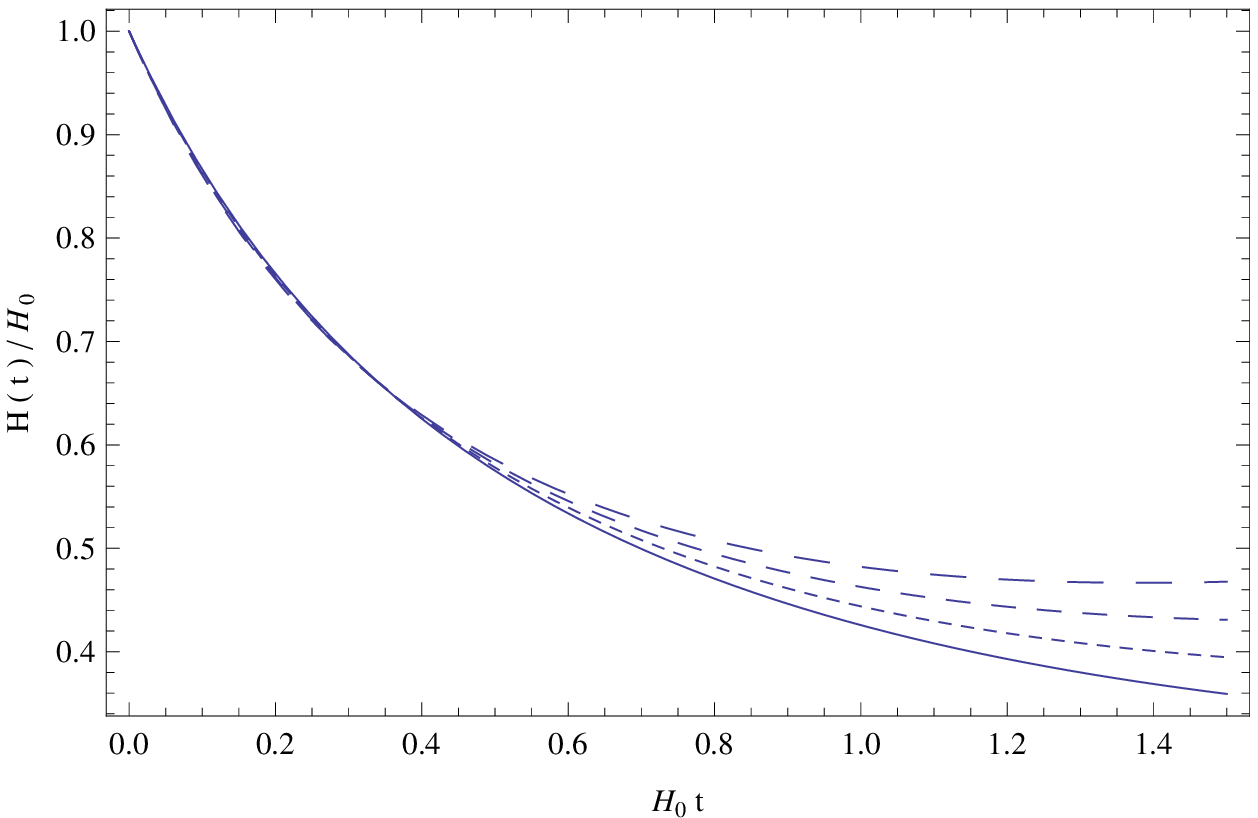}
\caption{ Time variation of the Hubble function $H(t)$ in the first Noether
symmetry solution of the generalized STT cosmological model for different
values of the parameter $\protect\chi $: $\protect\chi =0.15$ (solid curve),
$\protect\chi =0.25$ (dotted curve), $\protect\chi =0.35$ (dashed curve),
and $\protect\chi =0.45$ (long dashed curve), respectively.}
\label{fig2}
\end{figure}
\begin{figure}[h]
\centering
\includegraphics[width=0.45\textwidth]{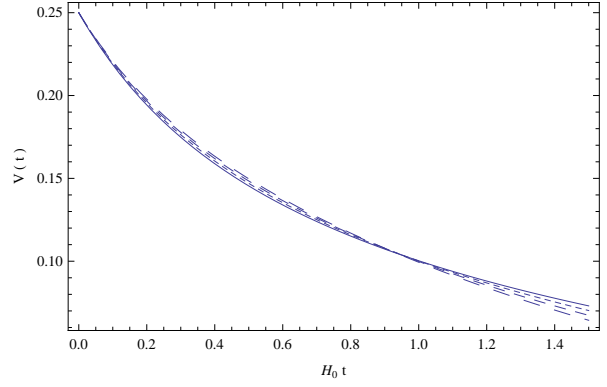}
\caption{ Time variation of the scalar field potential $V(t)$ in the first
Noether symmetry solution of the generalized STT cosmological model for
different values of the parameter $\protect\chi $: $\protect\chi =0.15$
(solid curve), $\protect\chi =0.25$ (dotted curve), $\protect\chi =0.35$
(dashed curve), and $\protect\chi =0.45$ (long dashed curve), respectively.
For the constant $V_0$ we have adopted the value $V_0=a_0^{3/2}/4n$.}
\label{fig3}
\end{figure}
\begin{figure}[h]
\centering
\includegraphics[width=0.45\textwidth]{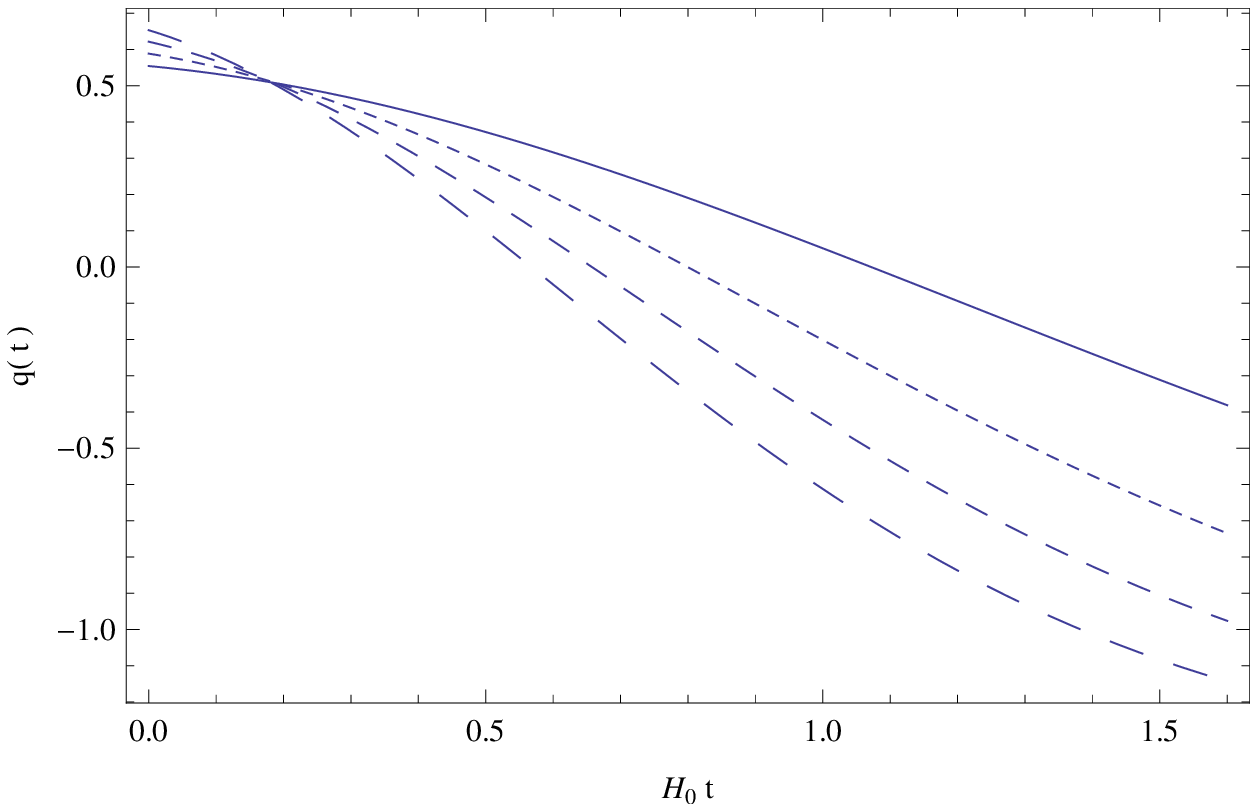}
\caption{ Time variation of the deceleration parameter $q(t)$ in the first
Noether symmetry solution of the generalized STT cosmological model for
different values of the parameter $\protect\chi $: $\protect\chi =0.15$
(solid curve), $\protect\chi =0.25$ (dotted curve), $\protect\chi =0.35$
(dashed curve), and $\protect\chi =0.45$ (long dashed curve), respectively.}
\label{fig4}
\end{figure}

The time variations of the scale factor, of the Hubble function, of the
scalar field potential and of the deceleration parameter for the first
Noether symmetry solution of the modified STT cosmological model are
represented in Figs.~\ref{fig1}-\ref{fig4} respectively

As one can see from Fig.~\ref{fig1}, the Universe described by the first
Noether symmetry solution of the generalized STT experiences an expansionary
phase, with the scale factor a monotonically increasing function of time.
The Hubble function, presented in Fig.~\ref{fig2}, is a monotonically
decreasing function of time in the early stages of cosmological evolution,
but, after reaching a minimum value, for large values of the parameter $\chi
$ it becomes a monotonically increasing function of time, thus indicating
the possibility of the reversal of the expansion to a collapsing phase. The
scalar field potential, depicted in Fig.~\ref{fig3}, is a monotonically
decreasing function of time, for all considered values of the parameter $%
\chi $. The deceleration parameter $q$, plotted in Fig.~\ref{fig4}, shows
that in the present model the Universe starts its evolution in a
decelerating phase, with $q>0$. However, in all cases the Universe enters in
an accelerating phase, with $q<0$, with the time spent in the decelerating
phase depending on the numerical value of $\chi $. For large values of $\chi
$, in the large time limit, the Universe ends its evolution with $q=-1.5$,
thus indicating that the present model allows super de Sitter accelerating
cosmological expansions.

\subsubsection{The constant of motion}

Since the constant of motion $\Sigma $ plays an important role in the study
of Noether symmetries, in the following we shall obtain the explicit form
for $\Sigma $. Using Eqs.~(\ref{i1}) and (\ref{i2}), we find for the
constant of motion $\Sigma $ given by%
\begin{equation}
\Sigma =\alpha \frac{\partial \mathcal{L}}{\partial \dot{a}}+\beta \frac{%
\partial \mathcal{L}}{\partial \dot{\phi}}.  \label{S1}
\end{equation}%
By substituting Eqs. (\ref{L}) and (\ref{xx}) into the Eqs.~(\ref{S1}), the
latter takes the form
\begin{equation}
\Sigma =-6Ka^{2}\dot{a}\phi ^{n}+3\left( 2nK-\frac{1}{n}\right) a^{3}\dot{%
\phi}\phi ^{n-1}.  \label{S2}
\end{equation}%
We rewrite Eq.~(\ref{S2}) as a Bernoulli differential equation for the
scalar field $\phi ^{n}$ in the form
\begin{equation}
\frac{d\phi ^{n}}{dt}=\frac{2Kn^{2}}{2Kn^{2}-1}\frac{d\ln a}{dt}\phi ^{n}+%
\frac{n^{2}\Sigma }{3\left( 2Kn^{2}-1\right) a^{3}}.  \label{S3}
\end{equation}%
The general solution of Eq. (\ref{S3}) is given by%
\begin{equation}
\phi ^{n}\left( a,t\right) =a^{\frac{2Kn^{2}}{2Kn^{2}-1}}\left[ \phi _{0}+%
\frac{n^{2}\Sigma }{3\left( 2Kn^{2}-1\right) }\int a^{\frac{8Kn^{2}-3}{%
1-2Kn^{2}}}dt\right] ,  \label{S4}
\end{equation}%
where $\phi _{0}$ is the arbitrary constant of integration. Since we have
already used $8Kn^{2}=3$ or equivalently $n_{\pm }=\pm \sqrt{\frac{3}{8K}}$,
thus Eq. (\ref{S4}) has a simple form

\begin{equation}
\phi ^{n}\left( a,t\right) =a^{-3}\left( \phi _{0}-\frac{4n^{2}\Sigma }{3}%
t\right) .  \label{S5}
\end{equation}%
Furthermore, if the constant of motion $\Sigma $ vanishes, then Eq. (\ref{S5}%
) will take a simpler form

\begin{equation}
\phi ^{n}\left( a\right) =\phi _{0}a^{-3}.  \label{S6}
\end{equation}%
The result given by Eq. (\ref{S6}) can be reobtained from the differential
equation $\frac{da}{\alpha }=\frac{d\phi }{\beta }$, where $\alpha \left(
a\right) =a$, $\beta \left( \phi \right) =-\frac{3}{n}\phi $.

\subsection{Cosmological solution 2}

Using the Noether symmetries presented in section III, we have found the
solutions given by Eqs. (\ref{s8})-(\ref{s10}). Here we recall these
solutions again%
\begin{equation}
F\left( \phi \right) =K\phi ^{n},V\left( \phi \right) =\lambda \phi
^{3c_{1}b_{3}},Z\left( \phi \right) =b_{0}\phi ^{n-2},  \label{A}
\end{equation}%
\begin{equation}
\alpha \left( a,\phi \right) =b_{1}a^{c_{1}}\phi ^{b_{4}},  \label{B}
\end{equation}%
\begin{equation}
\beta \left( a,\phi \right) =b_{2}a^{c_{1}-1}\phi ^{1+b_{4}},  \label{C}
\end{equation}%
where for simplicity we have denoted
\begin{equation*}
b_{0}=\frac{12Kn^{2}c_{1}\left(
1+c_{1}\right) }{\left( 1+2c_{1}\right) ^{2}}, b_{1}=\sqrt{\frac{2\left(
1+2c_{1}\right) c_{2}}{n\left( 1-c_{1}\right) }},
\end{equation*}
\begin{equation*}
b_{2}=-\frac{1}{c_{1}}%
\sqrt{\frac{2c_{2}}{1-c_{1}}\left( \frac{1+2c_{1}}{n}\right) ^{3}}, b_{3}=%
\frac{n}{1+2c_{1}},
\end{equation*}
\begin{equation*}
b_{4}=b_{3}\left( c_{1}^{2}-1\right),
\end{equation*}
  and $%
V_{0}=\lambda $. By inserting Eqs. (\ref{B}) and (\ref{C}) into the
differential equation $da/\alpha =d\phi /\beta $, thus integrating the
latter equation yields the invariant solution induced by the Noether
symmetry
\begin{equation}
\phi \left( a\right) =\phi _{1}a^{-\frac{1+2c_{1}}{nc_{1}}},  \label{k1}
\end{equation}%
where $\phi _{1}$ is an arbitrary constant of integration. By inserting Eqs.
(\ref{B}) and (\ref{C}) into Eqs. (\ref{w}) and (\ref{v}), we obtain the
solutions of Eqs.~(\ref{w}) and (\ref{v}) as given by%
\begin{equation}
w\left( a,\phi \right) =\phi a^{\frac{1+2c_{1}}{nc_{1}}},  \label{k2}
\end{equation}%
\begin{equation}
z\left( a,\phi \right) =\gamma ^{-1}a^{1-c_{1}}\phi ^{\frac{n(1-c_{1}^{2})}{%
1+2c_{1}}}.  \label{k3}
\end{equation}%
With the help of Eqs. (\ref{k2}) and (\ref{k3}), we obtain the scale factor $%
a$ and the scalar field $\phi $ given by
\begin{equation}
a\left( w,z\right) =\left[ \gamma zw^{\frac{n\left( c_{1}^{2}-1\right) }{%
1+2c_{1}}}\right] ^{\frac{c_{1}}{c_{1}-1}},  \label{k4}
\end{equation}%
\begin{equation}
\phi \left( w,z\right) =\left( \gamma z\right) ^{\frac{1+2c_{1}}{n\left(
1-c_{1}\right) }}w^{-c_{1}},  \label{k5}
\end{equation}%
respectively, where for simplicity we have denoted an arbitrary constant $%
\gamma $ as
\begin{equation}
\gamma =-\frac{1}{c_{1}}\sqrt{\frac{2c_{2}\left( 1-c_{1}\right)
\left( 1+2c_{1}\right) }{n}}.
\end{equation}
In order to have real functions $a\left(
w,z\right) $ and $\phi \left( w,z\right) $ as given by Eqs. (\ref{k4}) and (\ref%
{k5}), thus we impose the conditions $c_{1}<0$, and
\begin{equation}
\frac{%
2c_{2}\left( 1-c_{1}\right) \left( 1+2c_{1}\right) }{n}>0,
\end{equation}
 respectively. By inserting
Eqs.~(\ref{A}), (\ref{k4}) and (\ref{k5}) into the Lagrangian (\ref{L}), we
obtain the Lagrangian
\begin{equation}
\mathcal{L}=-\lambda w^{\frac{3nc_{1}}{1+2c_{1}}}-s_{0}\dot{w}\dot{z}w^{%
\frac{nc_{1}\left( 2+c_{1}\right) }{1+2c_{1}}-1},  \label{p}
\end{equation}%
expressed in the new variables $w$ and $z$, where we have denoted the
arbitrary constant $s_{0}$ as
\begin{equation}
s_{0}=6K\sqrt{\frac{2nc_{2}}{\left(
1-c_{1}\right) \left( 1+2c_{1}\right) }},
\end{equation}
and $c_{1}\neq $ $-1/2$. In the
following we shall present two cosmological solutions describing the
dynamics of the FRW Universe for the conditions $c_{1}\neq -2$ and $c_{1}=-2$
respectively.

\subsubsection{Specific case $c_{1}\neq -2$}

For the case $c_{1}\neq -2$, again by inserting Eq. (\ref{p}) into the
Euler-Lagrange Eqs.~(\ref{m2}) and (\ref{m3}), we obtain
\begin{equation}
\frac{d}{dt}\left( \dot{w}w^{\frac{nc_{1}\left( 2+c_{1}\right) }{1+2c_{1}}%
-1}\right) =0,  \label{u1}
\end{equation}%
and
\begin{eqnarray}
\ddot{z} &=&\frac{\lambda }{s_{0}}\left( \frac{3nc_{1}}{1+2c_{1}}\right) w^{%
\frac{nc_{1}\left( 1-c_{1}\right) }{1+2c_{1}}}=  \nonumber \\
&&\frac{\lambda }{s_{0}}\left( \frac{3nc_{1}}{1+2c_{1}}\right) \left(
w_{0}s_{1}t+s_{2}\right) ^{\frac{1-c_{1}}{2+c_{1}}},  \label{u2}
\end{eqnarray}%
respectively. By integrating Eqs.~(\ref{u1}) and (\ref{u2}) yields
\begin{equation}
w\left( t\right) =\left( w_{0}s_{1}t+s_{2}\right) ^{\frac{1+2c_{1}}{%
nc_{1}\left( 2+c_{1}\right) }},  \label{u3}
\end{equation}%
\begin{eqnarray}
z\left( t\right) &=&\frac{\lambda }{s_{0}}\left( \frac{3nc_{1}}{1+2c_{1}}%
\right) \int^{t}\int^{t_{2}}\left( w_{0}s_{1}t_{1}+s_{2}\right) ^{\frac{%
1-c_{1}}{2+c_{1}}}\times \nonumber \\
&&dt_{1}dt_{2}+
s_{3}t+s_{4},\frac{1-c_{1}}{2+c_{1}}\neq -1,  \label{u4}
\end{eqnarray}%
where we have used Eq.~(\ref{u3}) for obtaining Eq.~(\ref{u4}), and we have
denoted $s_{i}$, $i=1,2,3,4$ as the arbitrary constants of integration. We
have also denoted the arbitrary constant $w_{0}$ as $w_{0}=\frac{%
nc_{1}\left( 2+c_{1}\right) }{1+2c_{1}}$. By inserting Eq.~(\ref{u3}) into
Eq.~(\ref{u4}), then the latter takes the form
\begin{equation}
z\left( t\right) =\frac{\lambda z_{0}}{s_{0}}\left( w_{0}s_{1}t+s_{2}\right)
^{\frac{5+c_{1}}{2+c_{1}}}+s_{3}t+s_{4},  \label{u5}
\end{equation}%
where we have denoted the arbitrary constant $z_{0}$ as $z_{0}=\frac{1+2c_{1}%
}{\left( 5+c_{1}\right) nc_{1}s_{1}^{2}}$. By substituting Eqs. (\ref{u3})
and (\ref{u5}) into Eq.~(\ref{k4}), then we obtain the scale factor $a\left(
t\right) $ as
\begin{eqnarray}
a\left( t\right) &=&\left\{ \gamma \left[ \frac{\lambda z_{0}}{s_{0}}\left(
w_{0}s_{1}t+s_{2}\right) ^{\frac{5+c_{1}}{2+c_{1}}}+s_{3}t+s_{4}\right]
\right\} ^{\frac{c_{1}}{c_{1}-1}}\times  \nonumber \\
&&\left( w_{0}s_{1}t+s_{2}\right) ^{\frac{1+c_{1}}{2+c_{1}}}.  \label{u6}
\end{eqnarray}%
In order to simplify the obtained cosmological expressions we rescale the
time coordinate and the scale factor as $t\rightarrow t_{0}+t/w_{0}s_{1}$
and $a(t)\rightarrow a(t)\gamma ^{c_{1}/\left( c_{1}-1\right) }$,
respectively. We fix the constants $t_{0}$, $s_{1}$, $s_{2}$, $s_{3}$ and $%
s_{4}$ through the relations $w_{0}s_{1}t_{0}+s_{2}=0$, and $%
s_{3}t_{0}+s_{4}=0$, respectively. Then, by denoting
\begin{equation}
\frac{\lambda z_{0}}{s_{0}}=\frac{\sigma _{0}s_{3}}{w_{0}s_{1}}=\frac{s_{4}}{s_{2}}=\sigma ,
\end{equation}
and $m=3/\left( 2+c_{1}\right) $ respectively, we obtain first
\begin{eqnarray}
a(t) &=&\left( \sigma _{0}t^{m}+\sigma \right) ^{\frac{3-2m}{3(1-m)}}t^{%
\frac{6+m(m-6)}{3(1-m)}}=  \nonumber \\
&&a_{0}\left[ \left( \frac{t}{t_{0}}\right) ^{m}+1\right] ^{\frac{3-2m}{%
3(1-m)}}\left( \frac{t}{t_{0}}\right) ^{\frac{6+m(m-6)}{3(1-m)}},
\end{eqnarray}%
where $a_{0}=\sigma ^{\frac{3-2m}{3\left( 1-m\right) }}t_{0}^{\frac{6+m(m-6)%
}{3(1-m)}}$ and $t_{0}=\left( \sigma /\sigma _{0}\right) ^{1/m}$.

For the Hubble function $H(t)$, and for the deceleration parameter $q(t)$ we
obtain
\begin{equation}
H(t)=\frac{-m^{2}+[m(m+3)-6]\left( \frac{t}{t_{0}}\right) ^{m}+6m-6}{3(m-1)t%
\left[ \left( \frac{t}{t_{0}}\right) ^{m}+1\right] }.
\end{equation}%
and
\begin{eqnarray}
q(t) &=&3\left\{ m^{2}+\left[ 6-m(m+3)\right] \left( \frac{t}{t_{0}}\right)
^{m}-6m+6\right\} ^{-2} \times \nonumber\\
&&\Bigg\{ -m^{3}+7m^{2}+\Bigg[ m(m((5-2m)m+6)-21)\nonumber\\
&&+12\Bigg] \left( \frac{t}{%
t_{0}}\right) ^{m}+(m-1)[m(m+3)-6]\left( \frac{t}{t_{0}}\right)
^{2m}\nonumber\\
&&-12m+6\Bigg\} -1,
\end{eqnarray}
respectively.

By substituting Eqs. (\ref{u3}) and (\ref{u5}) into Eq. (\ref{k5}), then the
scalar field $\phi \left( t\right) $ is given by
\begin{equation}
\phi \left( t\right) =\phi _{0}\left[ \left( \frac{t}{t_{0}}\right) ^{m}+1%
\right] ^{\frac{2-m}{(m-1)n}}\left( \frac{t}{t_{0}}\right) ^{\frac{(m-2)^{2}%
}{(m-1)n}},  \label{u7}
\end{equation}%
where we have rescaled the scalar field as
\begin{equation}
\phi (t)\rightarrow \phi
(t)\gamma ^{\frac{1+2c_{1}}{n\left( 1-c_{1}\right) }},
\end{equation}
 and $\phi _{0}$ is
given by
\begin{equation}
\phi _{0}=\sigma ^{\frac{2-m}{(m-1)n}}t_{0}^{\frac{(m-2)^{2}}{%
(m-1)n}}.
\end{equation}

With the help of Eq. (\ref{u7}), the potential $V\left( t\right) $ can be
easily obtained from the relation $V\left( \phi \right) =\lambda \phi
^{3c_{1}b_{3}}$ as
\begin{equation}
V(\phi )=\lambda \phi ^{\frac{3nc_{1}}{1+2c_{1}}%
}=\lambda \phi ^{\frac{n\left( 2m-3\right) }{m-2}}.
\end{equation}

\begin{figure}[h]
\centering
\includegraphics[width=0.45\textwidth]{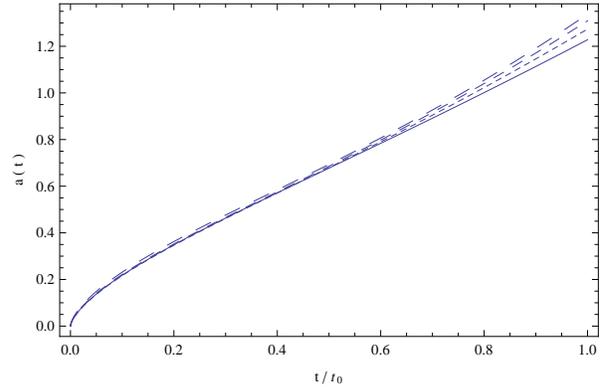}
\caption{ Time variation of the scale factor $a(t)$ in the second Noether
symmetry solution of the generalized STT cosmological model for different
values of the parameter $m$: $m=1.95$ (solid curve), $m =2.05$ (dotted
curve), $m =2.20$ (dashed curve), and $m =2.355$ (long dashed curve),
respectively. The value of the constant $a_0$ has been fixed as $a_0=1$. }
\label{fig5}
\end{figure}

The plots for the scale factor $a$, Hubble function $H$, deceleration
parameter $q$ and the potential $V$ are presented in Figs.~\ref{fig5}-\ref%
{fig8}, respectively.

\begin{figure}[h]
\centering
\includegraphics[width=0.45\textwidth]{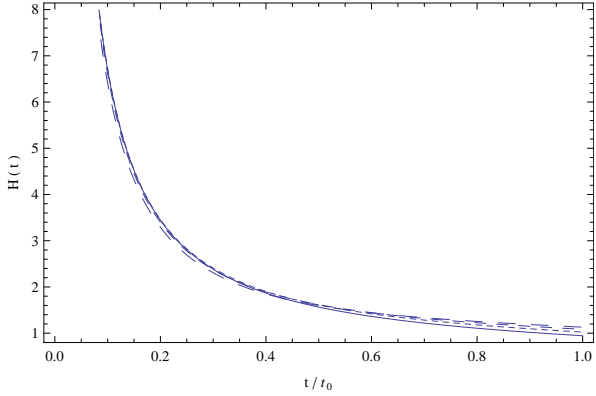}
\caption{ Time variation of the Hubble function $H(t)$ in the second Noether
symmetry solution of the generalized STT cosmological model for different
values of the parameter $m $: $m=1.95$ (solid curve), $m =2.05$ (dotted
curve), $m =2.20$ (dashed curve), and $m =2.355$ (long dashed curve),
respectively.}
\label{fig6}
\end{figure}

\begin{figure}[h]
\centering
\includegraphics[width=0.45\textwidth]{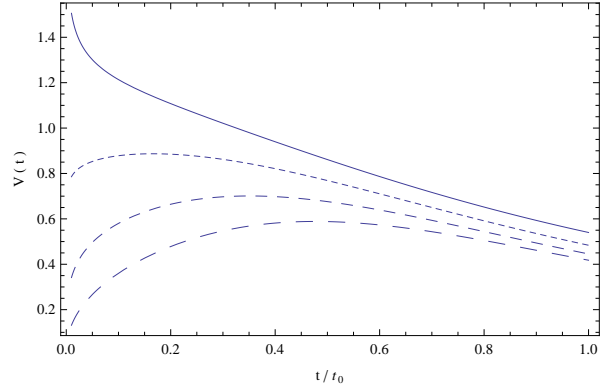}
\caption{ Time variation of the scalar field potential $V(t)$ in the second
Noether symmetry solution of the generalized STT cosmological model for
different values of the parameter $m $: $m=1.95$ (solid curve), $m =2.05$
(dotted curve), $m =2.20$ (dashed curve), and $m =2.355$ (long dashed
curve), respectively. The values of the integration constants have been
fixed so that $\protect\lambda \protect\phi _{0}^{2\left( 2m-3\right)
/\left( m-2\right) }=1$, while for $n$ we have adopted the value $n=2$.}
\label{fig7}
\end{figure}

\begin{figure}[h]
\centering
\includegraphics[width=0.45\textwidth]{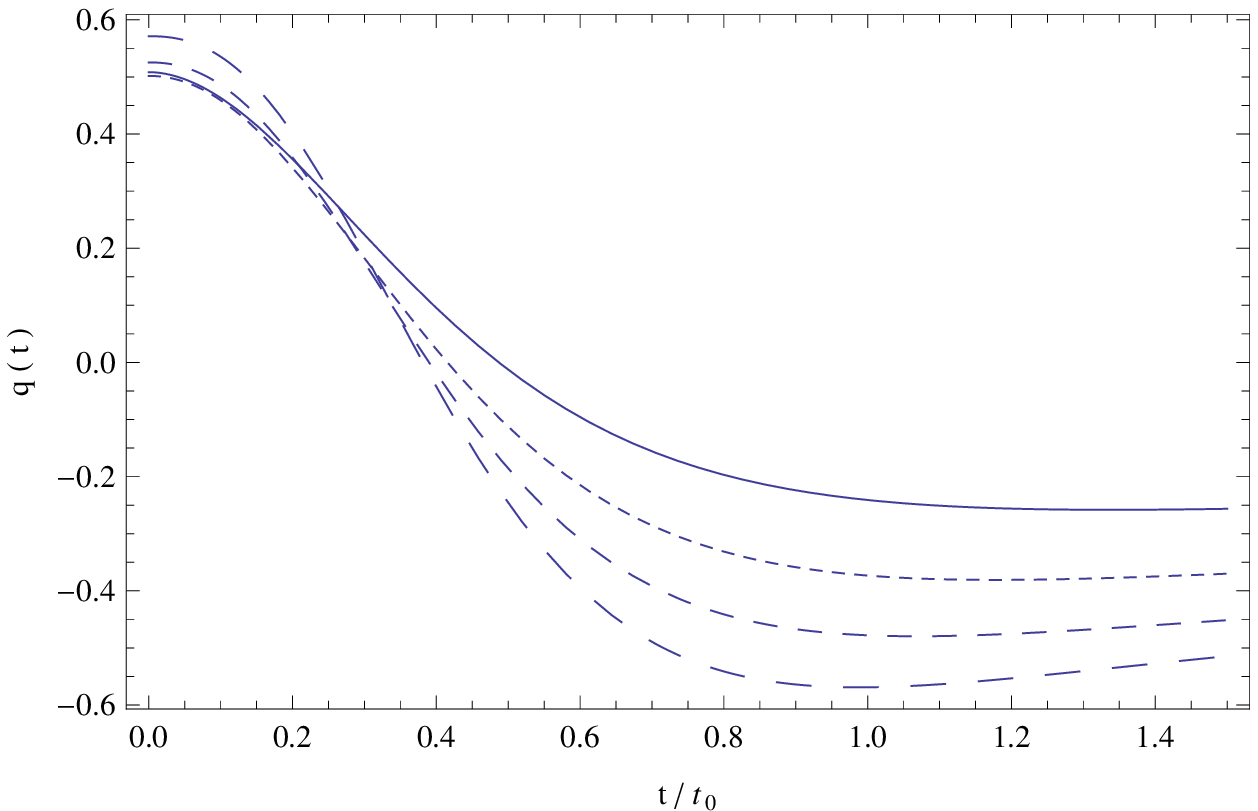}
\caption{ Time variation of the deceleration parameter $q(t)$ in the second
Noether symmetry solution of the generalized STT cosmological model for
different values of the parameter $m$: $m=1.95$ (solid curve), $m =2.05$
(dotted curve), $m =2.20$ (dashed curve), and $m =2.355$ (long dashed
curve), respectively.}
\label{fig8}
\end{figure}

The scale factor of this cosmological model, represented in Fig.~\ref{fig5},
shows that for the adopted values of the parameter $m$ the universes in a
state of expansion, with $a$ monotonically increasing in time. The Hubble
function, depicted in Fig.~\ref{fig6}, is a monotonically decreasing
function of time, tending, in the large time limit, to a constant value. The
scalar field potential, depicted in Fig.~\ref{fig7}, shows a complex
dynamical behavior. The potential is singular for $t\rightarrow 0$. Then it
decreases monotonically in time, reaches a minimum value, and begins again
to increase. The deceleration parameter, shown in Fig.~\ref{fig8}, indicates
the existence of a transition from a decelerating phase to an accelerating
one. It is important to point out that in the present model all relevant
cosmological quantities ($a$, $H$, $V$) are singular at the beginning of the
cosmological evolution corresponding to $t=0 $.

In the next Section, we shall present the cosmological solution for the
particular case $c_{1}=-2$.

\subsubsection{Specific case $c_{1}=-2$, $c_{2}<0$}

For the case $c_{1}=-2$, then the Lagrangian (\ref{p}) takes the simple form
\begin{equation}
\mathcal{L}=-\lambda w^{2n}-m_{0}\frac{\dot{w}}{w}\dot{z},  \label{b1}
\end{equation}%
where we have denoted the arbitrary constant $m_{0}$ as $m_{0}=2K\sqrt{%
-2nc_{2}}$. Again by inserting Eq. (\ref{b1}) into Euler-Lagrange Eqs. (\ref%
{m2}) and (\ref{m3}), we obtain
\begin{equation}
\frac{d}{dt}\left( \frac{\dot{w}}{w}\right) =0,  \label{z1}
\end{equation}%
\begin{equation}
\ddot{z}=\frac{2n\lambda }{m_{0}}w^{2n}.  \label{z2}
\end{equation}%
respectively. By integrating Eqs. (\ref{z1}) and (\ref{z2}), we find
\begin{equation}
w\left( t\right) =h_{2}e^{h_{1}t},  \label{z3}
\end{equation}
\begin{equation}
z\left( t\right) =z_{1}e^{2nh_{1}t}+h_{3}t+h_{4},  \label{z4}
\end{equation}%
where $h_{i}$, $i=1,2,3,4$ are the arbitrary constants of integration, and
we have denoted the arbitrary constant $z_{1}$ as $z_{1}=\frac{\lambda
h_{2}^{2n}}{2nm_{0}h_{1}^{2}}$. By substituting Eqs.~(\ref{z3}) and (\ref{z4}%
) into Eq.~(\ref{k4}), then we obtain the scale factor $a\left( t\right) $
given by%
\begin{equation}
a(t)=A_{0}\left( z_{1}e^{2nh_{1}t}+h_{3}t+h_{4}\right) ^{2/3}e^{-2nh_{1}t/3},
\label{z5}
\end{equation}%
where $A_{0}=\left[ 3h_{2}^{-n}\sqrt{-2c_{2}/n}/2\right] ^{2/3}$. Without
any loss of generality we can take for the constant $h_{4}$ the value zero, $%
h_{4}=0$. For the Hubble function we find
\begin{equation}
H(t)=\frac{2}{3}\left[ \frac{h_{3}\left( 1-2h_{1}nt\right) }{%
z_{1}e^{2h_{1}nt}+h_{3}t}+h_{1}n\right] .
\end{equation}%
At the initial moment $t=0$ we have $a(0)=a_{0}=A_{0}z_{1}^{2/3}$, and $%
H(0)=H_{0}=(2/3)\left( h_{1}n+h_{3}/z_{1}\right) $, which gives $%
h_{3}/z_{1}=3H_{0}/2-h_{1}n$. Hence, by denoting $h_{0}=nh_{1}$ and $\xi
=3H_{0}/2h_{0}$, we can write the scale factor and the Hubble function as
\begin{equation}
a(t)=a_{0}\left[ e^{2h_{0}t}+\left( \xi -1\right) h_{0}t\right]
^{2/3}e^{-2h_{0}t/3},
\end{equation}%
\begin{equation}
H(t)=\frac{2}{3}h_{0}\left[ \frac{\left( \xi -1\right) \left(
1-2h_{0}t\right) }{e^{2h_{0}t}+\left( \xi -1\right) h_{0}t}+1\right] ,
\end{equation}%
respectively. For the deceleration parameter $q(t)$, we obtain
\begin{equation}
q(t)=\frac{3(\xi -1)\left[ 4e^{2h_{0}t}(1-h_{0}t)+\xi -1\right] }{2\left[
(1-\xi )h_{0}t+e^{2h_{0}t}+\xi -1\right] ^{2}}-1.
\end{equation}%
By substituting Eqs.~(\ref{z3}) and (\ref{z4}) into Eq.~(\ref{k5}), then we
obtain the scalar field
\begin{equation}
\phi \left( t\right) =h_{2}^{2}e^{2h_{1}t}\left[ \frac{3}{2}\sqrt{\frac{%
-2c_{2}}{n}}\left( z_{1}e^{2nh_{1}t}+h_{3}t+h_{4}\right) \right] ^{-1/n}.
\label{z6}
\end{equation}%
With the help of Eq. (\ref{z6}), since we have used $c_{1}=-2$, the
potential $V\left( t\right) $ can be easily obtained from the relation
\begin{equation}
V\left( \phi \right) =\lambda \phi ^{3c_{1}b_{3}}=\lambda \phi ^{\frac{%
3nc_{1}}{1+2c_{1}}}=\lambda \phi ^{2n},
\end{equation}
 as
\begin{equation}
V(t)=V_{1}e^{4h_{0}t}\left[ e^{2h_{0}t}+\left( \xi -1\right) h_{0}t\right]
^{-2},
\end{equation}%
where $V_{1}=-2n\lambda h_{2}^{4n}/9c_{2}z_{1}^{2}$.

\begin{figure}[h]
\centering
\includegraphics[width=0.45\textwidth]{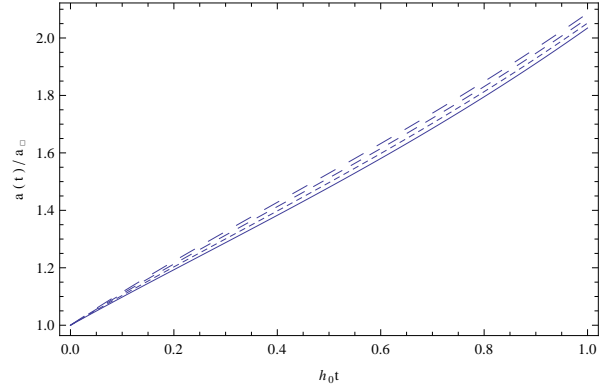}
\caption{ Time variation of the scale factor $a(t)$ in the third Noether
symmetry solution of the generalized STT cosmological model for different
values of the parameter $\protect\xi$: $\protect\xi=1.5$ (solid curve), $%
\protect\xi =1.6$ (dotted curve), $\protect\xi =1.7$ (dashed curve), and $%
\protect\xi =1.8$ (long dashed curve), respectively. }
\label{fig9}
\end{figure}

The time variations of the scale factor, of the Hubble function, of the
scalar field potential and of the deceleration parameter for the third
Noether symmetry cosmological solution are presented in Figs.~\ref{fig9}-\ref%
{fig12} respectively.

\begin{figure}[h]
\centering
\includegraphics[width=0.45\textwidth]{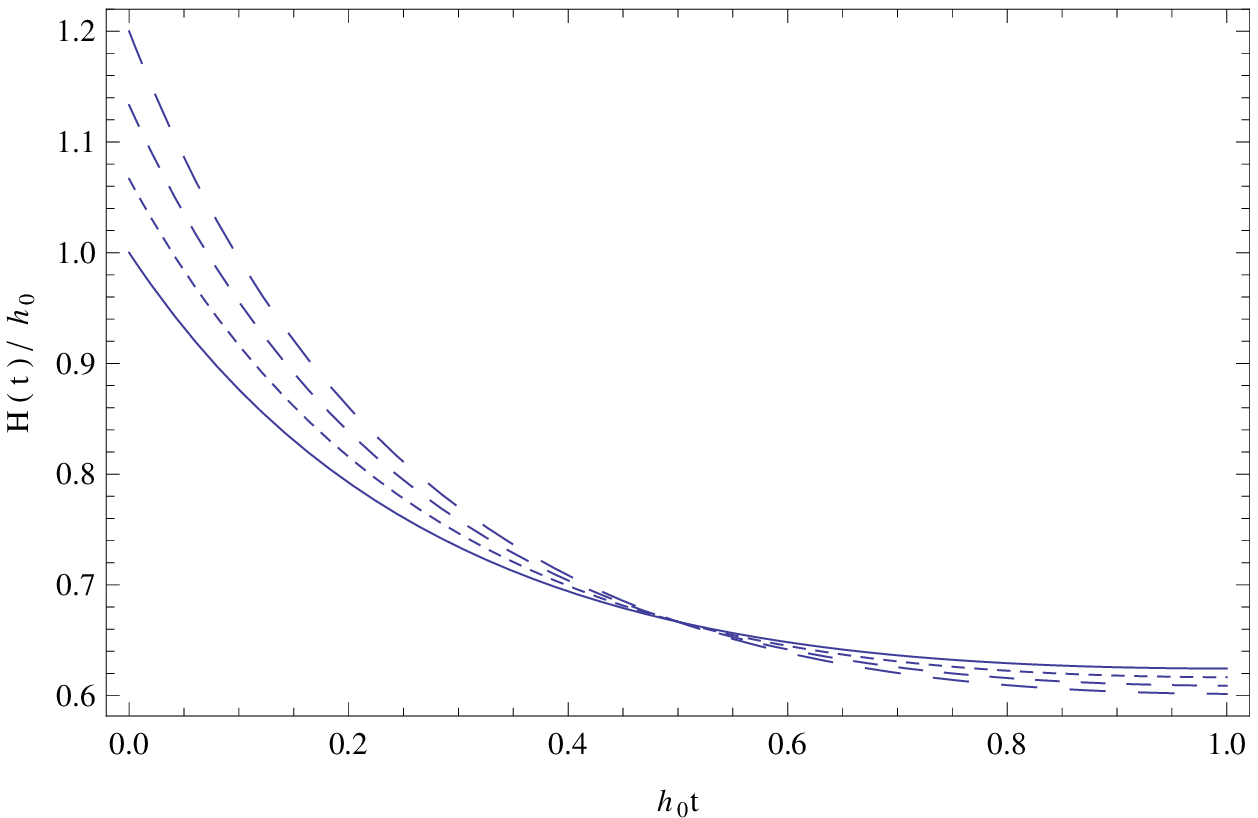}
\caption{ Time variation of the Hubble function $H(t)$ in the third Noether
symmetry solution of the generalized STT cosmological model for different
values of the parameter $\protect\xi$: $\protect\xi=1.5$ (solid curve), $%
\protect\xi =1.6$ (dotted curve), $\protect\xi =1.7$ (dashed curve), and $%
\protect\xi =1.8$ (long dashed curve), respectively.}
\label{fig10}
\end{figure}

\begin{figure}[h]
\centering
\includegraphics[width=0.45\textwidth]{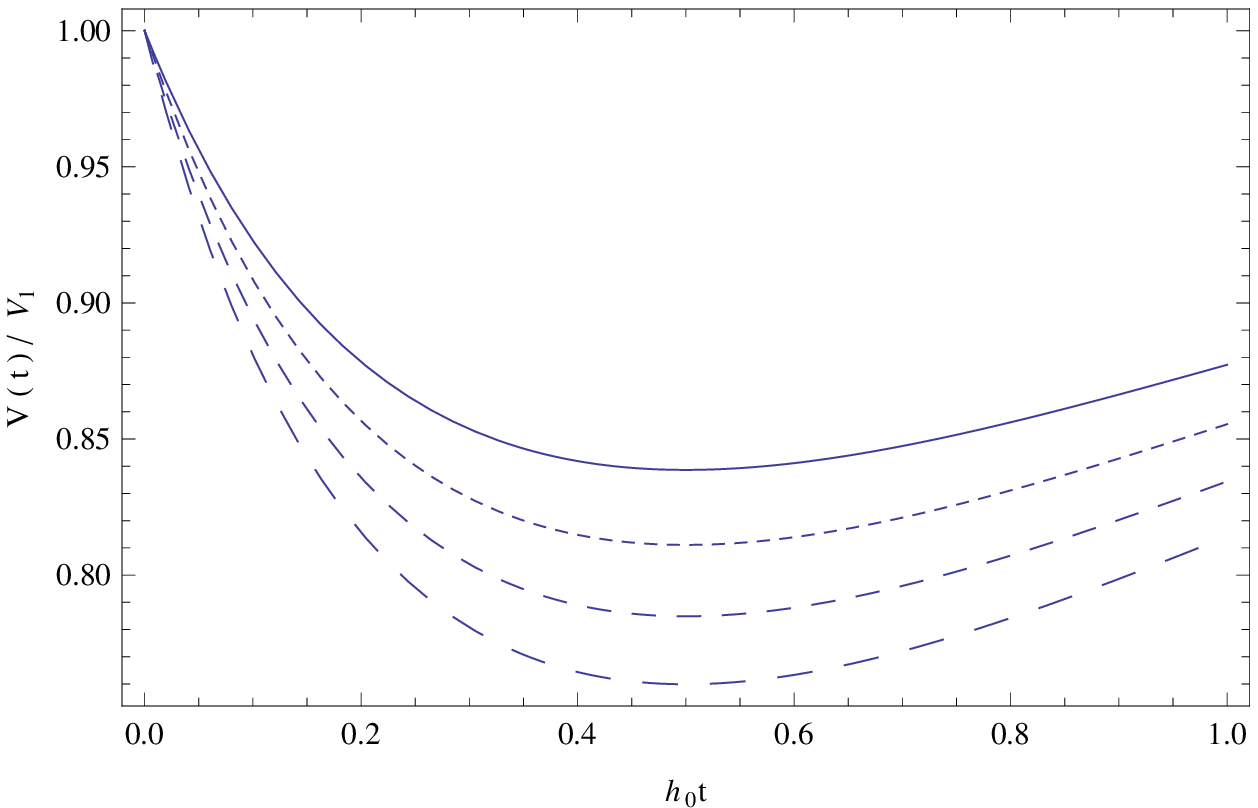}
\caption{ Time variation of the scalar field potential $V(t)$ in the third
Noether symmetry solution of the generalized STT cosmological model for
different values of the parameter $\protect\xi$: $\protect\xi=1.5$ (solid
curve), $\protect\xi =1.6$ (dotted curve), $\protect\xi =1.7$ (dashed
curve), and $\protect\xi =1.8$ (long dashed curve), respectively. }
\label{fig11}
\end{figure}

\begin{figure}[h]
\centering
\includegraphics[width=0.45\textwidth]{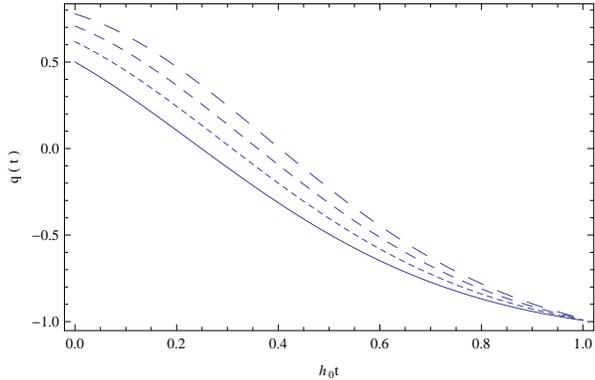}
\caption{ Time variation of the deceleration parameter $q(t)$ in the third
Noether symmetry solution of the generalized STT cosmological model for
different values of the parameter $\protect\xi$: $\protect\xi=1.5$ (solid
curve), $\protect\xi =1.6$ (dotted curve), $\protect\xi =1.7$ (dashed
curve), and $\protect\xi =1.8$ (long dashed curve), respectively.}
\label{fig12}
\end{figure}

Similarly to the behavior of the previous cosmological solutions, the scale
factor, depicted in Fig~\ref{fig9} is a monotonically increasing function of
time. The Hubble function, shown in Fig.~\ref{fig10}, monotonically
decreases during the expansion of the Universe, from its initial value $H_0$%
. The scalar field potential, represented in Fig.~\ref{fig11}, indicates a
complex behavior of the scalar field. After a period of monotonic decrease,
the field potential reaches a minimum value, and after that it starts
increasing. The deceleration parameter, presented in Fig.~\ref{fig12},
indicates that the evolution of the cosmological model starts in a
decelerating phase, with $q>0$. Then the Universe enters into an
accelerating phase, and in the large time limit the evolution becomes of de
Sitter type, with $q\rightarrow -1$.

\section{Conclusions and final remarks}

\label{sect4}

In the present paper we have investigated, by using Noether symmetry
techniques, the cosmological solutions of a generalized Brans-Dicke type STT
gravitational model, in which there is a supplementary coupling between the
scalar field and its kinetic energy term. From a mathematical point of view,
we have obtained three analytical solutions satisfying the condition $L_{%
\mathbf{X}}\mathcal{L}=\mathbf{X}\mathcal{L}=0$, where $\mathcal{L}$ is the
cosmological Lagrangian. Next by introducing two new variables $z$ and $w$,
defined by $z=z(a,\phi )$ and $w=w(a,\phi )$, and by fixing the numerical
values of the arbitrary integration constants, we have obtained three
cosmological solutions describing the dynamics of an FRW Universe. The
cosmological properties of the solutions have been investigated in detail.
The main, and common characteristic of these solutions is that they all
describe the transition from decelerating to accelerating phases. Another
interesting feature of the solutions is related to the complex time behavior
of the scalar field potential, which for some numerical values of the model
parameters can either increase to a maximum value, decreasing afterwards, as
is the case in the second model (see Fig.~\ref{fig7}), or decreases from a
maximum to a minimum value, from which it starts to increase again (model
three, Fig.~\ref{fig11}). Of course, due to the presence of some arbitrary
numerical parameters the cosmological dynamics is strongly dependent on
their numerical values. In the present approach we have fixed these unknown
parameters from the initial conditions of the cosmological expansion, fixed
at an arbitrary moment $t=0$. But of course the exact numerical values of
these parameters must be obtained from the full confrontation of the models
with the observational results.

The Planck satellite has recently provided high-quality data  of the CMB
anisotropies  \citep{Pl1,Pl2,Pl3}. These data have open new perspectives for the study of the physical processes that dominated  the cosmological evolution of the  very early Universe. An important result that can be inferred from the observational data is that they support the inflationary paradigm \citep{Guth} as an explanation for the evolution of the primordial phase in the existence of the Universe. Hence, from observational point of view it turns out that presently the full set of cosmological observations can be explained by adopting a
minimal theoretical setup, in which inflation is driven by a single scalar field $\phi $ (the inflaton),  minimally coupled to gravity. The scalar field has a canonical kinetic term $\dot{\phi}^2/2$, and, moreover,  its time evolution is determined by a flat potential $V(\phi)$ in the slow roll
phase. Since in the time of inflation the energy of the physical processes is very high, there is no firm knowledge of the physics of particles and fields during inflation. That's why a large variety of scalar field potentials have been proposed so far in the literature, and the determination of the possible forms of $V(\phi)$ is a central issue of research in present day cosmology and theoretical physics. On the other hand the possibility of non-minimal couplings between the inflation field and gravity, like, for example, those introduced in Eq.~(\ref{Act}) cannot be rejected a priori. In this context, the imposition of the Noether symmetry, which could be interpreted as a fundamental symmetry of nature, can provide a powerful method for the determination of the inflaton field potential. We have obtained the explicit form of several power-law type potentials, which could be relevant for the study of the inflationary epoch. In order to evaluate the cosmologically relevant parameters during inflation, the time evolution of the scalar
field can be expressed in terms of the e-fold number $N$, given by $N\equiv \ln \left(a_e/a\right)$, where $a_e$ is the value of the scale factor at the end of the inflation. $N$ is related to the scalar field potential by the relation $N=\left(1/M_P^2\right)\int _{\phi _e}^{\phi}{\left[V(\phi)/V'(\phi)\right]d\phi}$ \citep{Liddle}. Hence, if an inflationary potential is given, one can find first find $\phi (N)$. Once $\phi (N)$ is known, one can immediately obtain another important cosmological parameters, like the scalar spectral index, and the tensor to-scalar
ratio, respectively. Therefore, once these parameters have been computed, one can compare the predictions of the inflationary
potentials obtained via Noether symmetry with the Planck 2015 results. Such a comparison may also fix the numerical values of the parameters (constants of integration) of the Noether potentials. Hence for all inflationary type Noether models non trivial constraints on the parameters can be obtained from the CMB data.

To conclude, in the present paper we have presented some
theoretical tools that could help in the mathematical analysis of the
scalar-tensor theories, and in the investigation of their cosmological
implications.

\section*{Acknowledgments}

We would like to thank to the anonymous referee for comments and suggestions that helped us to improve our manuscript. T.H. would like to thank the Yat Sen School and the Department of Physics of the Sun Yat Sen
University in Guangzhou, P. R. China,  for the kind hospitality offered during the
preparation of this work.

\end{document}